\documentclass[sigchi]{acmart}

\usepackage{booktabs} 
\usepackage[disable]{todonotes}

\usepackage{balance}

\copyrightyear{2019} 
\acmYear{2019} 
\setcopyright{acmcopyright}
\acmConference[CHI 2019]{CHI Conference on Human Factors in Computing Systems Proceedings}{May 4--9, 2019}{Glasgow, Scotland Uk}
\acmBooktitle{CHI Conference on Human Factors in Computing Systems Proceedings (CHI 2019), May 4--9, 2019, Glasgow, Scotland Uk}
\acmPrice{15.00}
\acmDOI{10.1145/3290605.3300605}
\acmISBN{978-1-4503-5970-2/19/05}

\newcommand{\ken}[1]{\todo[color=yellow, inline]{\textbf{Kenny: } #1}}

\newcommand{\name}{{\em Empath-D\ }}
\newcommand{\names}{{\em Empath-D}}

\begin{document}
\title[Augmented Virtuality Impairment Simulation for Accessibility Design]{Examining Augmented Virtuality Impairment Simulation for Mobile App Accessibility Design}

\author{Kenny Tsu Wei Choo}
\affiliation{%
\institution{Singapore Management University}
}
\email{kennychoo@smu.edu.sg}

\author{Rajesh Krishna Balan}
\affiliation{%
\institution{Singapore Management University}
}
\email{rajesh@smu.edu.sg}

\author{Youngki Lee}
\affiliation{%
\institution{Seoul National University}
}
\email{youngkilee@snu.ac.kr}

\renewcommand{\shortauthors}{K.T.W. Choo et al.}

\begin{abstract}
With mobile apps rapidly permeating all aspects of daily living with use by all segments of the population, it is crucial to support the evaluation of app usability for specific impaired users to improve app accessibility. 
In this work, we examine the effects of using our \textit{augmented virtuality} impairment simulation system--\names--to support experienced designer-developers to redesign a mockup of commonly used mobile application for cataract-impaired users, comparing this with existing tools that aid designing for accessibility.
We show that the use of augmented virtuality for assessing usability supports enhanced usability challenge identification, finding more defects and doing so more accurately than with existing methods. 
Through our user interviews, we also show that augmented virtuality impairment simulation supports realistic interaction and evaluation to provide a concrete understanding over the usability challenges that impaired users face, and complements the existing guidelines-based approaches meant for general accessibility.


\end{abstract}


%
%




\maketitle

\section{Introduction}
\ken{REBUTTAL: We will address the flow, omissions, minor spelling and grammar mistakes as indicated by R1, R2 and 2AC in our revision.}

As we well know, mobile apps are a primary source of interaction for a large percentage of the population. These apps are usually developed for use by a younger and generally able-bodied population. However, the global elderly population is increasing and is expected to reach 16.7\% by 2050~\cite{he_aging_2016}. How usable would current mobile apps be when used by an elderly population with diminished eyesight, hearing, and other senses?

In this paper, we present the results of a study designed to answer that question as it relates to a \textit{visual disability} of cataracts and from a mobile app developer's perspective. In particular, how can we provide an effective development environment that allows a mobile app developer to refactor the user interface components of their app to make it more accessible when used by an elderly population.

\begin{figure*}[tbh]
\includegraphics[width=0.95\textwidth]{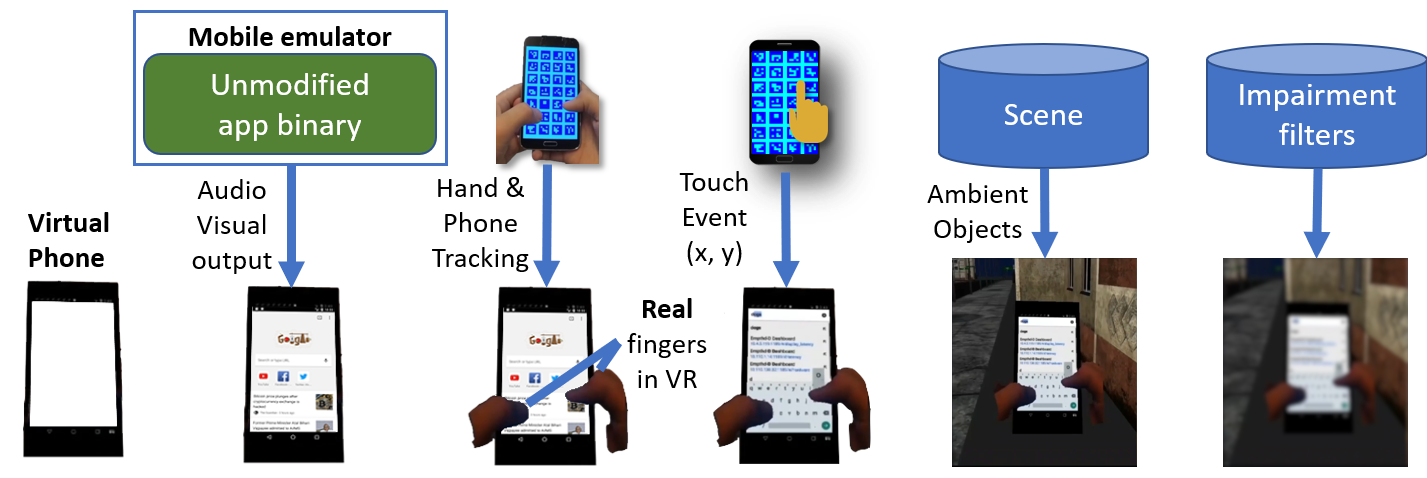}
\caption{Composing the augmented virtuality view in \name}
\label{fig:composing_views_empathd}
\Description[A series of six images and accompanying text showing a step-by-step composition of the augmented virtuality view in Empath-D]{A series of six images and accompanying text showing a step-by-step composition of the augmented virtuality view in Empath-D. The first image shows a blank virtual phone. The second image shows the mobile emulator running an unmodified app binary, which forwards the audio visual output to the virtual phone. The third image shows the physical phone with its arrays of fiduciary markers and dark blue on light blue backgrounds that are used for hand and phone tracking. Real fingers are shown on top of the virtual phone. The fourth image shows the forwarding of the touch events from the physical to virtual phone. The fifth image shows the rendering of the chosen scene into the VR. The last image shows the application of the impairment filters.}
\end{figure*}

To provide this perspective, we used \name~\cite{kim_empath-d:_2018}, our system that was developed and demonstrated at MobiSys 2018 as it provides a novel Virtual Reality (VR) environment where developers can experience how their mobile apps would operate when used by individuals with various visual physical ailments such as cataracts and glaucoma--both common ailments amongst an elderly population.
\name has many novel elements such as 1) its use of virtual reality allows many different visual and auditory ailments to be experienced, and 2) it uses a real phone device held by the user to capture actual touchscreen events and these events are replayed in the VR world on a virtual phone that is running the actual app in an Android emulator. 
Thus the user can run the actual app in the VR environment and interact with it naturally--providing an \textit{augmented virtuality} experience~\cite{milgram1994taxonomy}. Figure~\ref{fig:composing_views_empathd} provides a visual description of how \name works.

We conducted a rigorous design study to understand how developers could use a solution such as \name to build apps that were more accessible to a disadvantaged population. 
In particular, without any significant loss of generality, we focused on an elderly population suffering from cataracts. 
To conduct this study, we first recruited 4 elderly (>60 years old) participants suffering from cataracts to understand the usability challenges they faced when using a popular mobile app--in this case, we used the {\em Instagram} app on Android. 
We then recruited 10 app developers, between the ages of  21 and 31, to modify a mockup of Instagram to be more accessible. 
We compared their performance with and without \name and evaluated their output against both accessibility guidelines and subjective evaluation by our 4 elderly participants.

Our work yields insights on a number of issues involved in designing mobile apps with an augmented virtuality impairment simulator interface like \name and over existing tools:

\begin{itemize}
    \item How do existing accessibility checking and inspection tools perform in a design task?
    \item What are the usability challenges that an augmented virtuality impairment simulation identify that existing solutions cannot?
    \item What are the challenges with using augmented virtuality simulation for design?
    \item Is an augmented virtuality impairment simulator more usable for design?
\end{itemize}

Overall, our analysis of these questions contributes to a fundamental body of knowledge on the use of augmented virtuality to design for impairments that is novel and has not been adequately addressed.


\section{Background}

\ken{REBUTTAL: We thank 1AC for the references to help contextualise our findings. We believe that Silverman's work will help to provide an interesting perspective into disability simulation, and Lazar's work together with a stronger description of [13, 24] will provide the insights on the problems of existing approaches. We will incorporate this in our revision.}

\subsection{Mobile Application Accessibility}
While much work has been done in the area of accessibility for the web~\cite{caldwell_web_2008, power_guidelines_2012, clegg2014investigating, lazar_improving_2004}, considerably less work has been done on mobile app accessibility. With web content increasingly being accessed from mobile devices, the Worldwide Web Consortium Web Accessibility Initiative (W3C WAI) recognised the convergence and released a note to help map the Web Content Accessibility Guidelines (WCAG)~\cite{caldwell_web_2008} for mobile devices~\cite{patch_mobile_nodate}.
Beyond web accessibility, past work has characterised mobile application accessibility in areas such as mobile health apps~\cite{milne_accessibility_2014}, for the blind~\cite{carvalho_accessibility_2018}, or to develop frameworks for analysis at scale~\cite{ross_epidemiology_2017}. 

While the WCAG has been in place since 1999 (WCAG 1.0~\cite{chisholm_web_nodate}, and subsequent updates released to its current version of WCAG 2.1), large percentages of web sites are still inaccessible to users with disabilities~\cite{lopes_web_2010, davis_most_2017}. 
Lazar et al.~\cite{lazar_improving_2004} provide some insights on the challenges of implementing accessibility on webpages from developers' perspectives. 
The most pertinent perspectives to our work, were web developers responses that they \textit{lack training}, \textit{had inadequate software tools, and were confused by the accessibility guidelines}.
Our work developing \name~\cite{kim_empath-d:_2018} and validation in design here aim to address these.
Our focus, is on \textit{tools that support accessibility design for mobile apps}. 

To support developers to consider accessibility in their designs, Apple~\cite{noauthor_ios_nodate} and Google~\cite{noauthor_android_nodate} both released mobile accessibility guidelines, which are considerably simpler than the WCAG. 
They also developed tools--Google Accessibility Scanner (GAS)~\cite{noauthor_accessibility_nodate} and Apple Accessibility Inspector~\cite{noauthor_apple_2015}--that examine apps for accessibility on a screen-by-screen basis. 
We incorporated Google Accessibility Scanner in our baseline condition as it provides the most direct assessment of Android apps for accessibility considerations.

\subsection{Virtual Impairment Simulation for Design}

Virtual impairment simulation as an approach to support iterative design may offer significant advantages by offering authentic first-person perspectives. 
However, this needs to be considered in the frame of the adaptability of disabled users.
Silverman et al.~\cite{silverman_stumbling_2015} note that many impairment simulations often emphasise the initial challenges of becoming disabled.
This may lead to overcompensation for the abilities of disabled users in designs, which may not sit well with users who have adapted around their disabilities.
Our vision in developing \name~\cite{kim_empath-d:_2018} is to provide a flexible system that may be calibrated to suit the desired design objectives for target user abilities.
The detailed validation studies in this paper demonstrate the viability of \name as an approach to target user abilities (in particular, visual abilities).
This provides a customisable platform by which future work in the research community may implement impairment simulations that are accurate to the \textit{adapted abilities} of impaired users and examine the design outcomes.

Virtual impairment simulation for design has also been employed in multiple settings. 
In one of the earlier works on the subject, Higuchi et al.~\cite{higuchi_simulating_1999} proposed a tool to simulate the visual capabilities of the elderly on digital scans of photos of control panels.
Mankoff et al.~\cite{mankoff_evaluating_2005} developed a tool to simulate visual and motor impairments for any desktop application (including web interfaces). 
More recently, a head mounted display with visual impairment simulation was used in a virtual 3D model of a city to understand the issues involved in wayfinding and target location~\cite{vayrynen_head_2016}. 
Ates et al.~\cite{ates_immersive_2015} implemented an augmented reality see-through approach--using stereoscopic cameras mounted on an Oculus VR--to simulate impairments to inspect mobile applications. 
While their~\cite{vayrynen_head_2016, ates_immersive_2015} work is the closest to our tool \name~\cite{kim_empath-d:_2018} that is used in our studies--our goal was to examine the usability challenges that may be identified and improved in mobile apps using an augmented virtuality approach in performing actual mobile application design.

\subsection{Augmented Virtuality Impairment Simulation}
We used our augmented virtuality system, \names~\cite{kim_empath-d:_2018} to conduct our studies to observe and investigate the effects of using augmented virtuality impairment simulation in designing mobile apps. 

\name is an augmented virtuality system composed of a VR display mounted with a depth camera, a dummy phone, with a mobile emulator running on a computer. 
The software runs off the same computer and tightly orchestrates the different pieces of hardware such as to provide sufficiently real-time interaction performance~\cite{kim_empath-d:_2018} for most purposes (237.7 msec end-to-end touch latency). 
To experience \names, an unmodified app binary--the designed app--is run in the emulator (See Figure~\ref{fig:composing_views_empathd}). 
The RGBD images from the depth camera and the fiduciary marker array on the dummy phone are processed to support hand and phone tracking. 
Touch events on the dummy phone are also forwarded and processed. 
Finally, the scene and impairments are rendered--with the corresponding tracking--onto the VR display using the graphics engine.

To use Empath-D, the designer-developers simply need to 1) provide an unmodified app apk to be tested to the Empath-D simulator, 2) configure the impairment conditions and severity to test, and 3) wear a head-mounted VR device and use the application in the simulated environment to identify accessibility problems.

\section{Study Method}
We designed and conducted two studies to examine if \name is a useful tool to design mobile applications for impairment-specific accessibility. 
In the studies, we focused on 1) cataract-impaired users, one of the most common vision problems that the elderly experience and 2) a social media application--Instagram--an archetype that encapsulates many common interactions in mobile apps.

In the first study (S1), we recruited cataract-impaired users to examine the usability challenges that they faced under a few everyday use cases of Instagram. In the second study (S2), we recruited experienced designer-developers and got them to redesign a mock-up of Instagram using \name and compared that against Google Accessibility Scanner.

\subsection{Study 1: Cataract Impaired Users}
\subsubsection{Participants}
We recruited 4 (all female), ages 65-71 cataract-impaired users (self-reported to be verified by doctors) to identify the usability challenges that they experience in using Instagram. 
We used strict participant selection criteria to reduce the effects of other visual impairments on the study condition (cataracts). 
In particular, all participants were selected for mild-moderate cataracts in \textit{both} eyes, with minimal or no other visual impairments (e.g., low-degree myopia, no glaucoma, no age-related macular degeneration).
With cataracts being a predominantly age-related disease~\cite{foster_risk_2003}, all participants unavoidably also had presbyopia--the loss of elasticity in the lens of the eye, which causes issues with near focus. 

To confirm that our selection criteria was effective and that the usability challenges identified by participants were based on a similar level of cataract impairment, we asked participants to complete the CatQuest-9SF questionnaire~\cite{lundstrom_catquest-9sf_2009}. 
The CatQuest-9SF is a questionnaire originally developed to measure pre- and post-cataract surgery outcomes.
It was constructed in Swedish (with an English translated version) but has been translated and examined (under Rasch analysis) for Malay, Chinese, and Italian populations, and has consistently shown good psychometric properties~\cite{ adnan_catquest-9sf_2018, skiadaresi_italian_2016}.

\subsubsection{Method}

Participants came in for a 1 hour session, where we adopted a master-apprentice frame to explore their use of Instagram on a Samsung Galaxy S7. 
To identify the usability issues they faced with Instagram, we used an unmodified version with default settings and no accessibility options enabled. 
They were asked to 1) use the phone with or without corrective glasses as they would normally (in line with the notion of everyday living from~\cite{lundstrom_catquest-9sf_2009}), and 2) hold their phones at a normal distance of ~25-30cm from their eyes, and not compensate for visual problems (e.g., holding phone very close to see).
We focused the participants on three particular use scenarios (T1-T3, See Table~\ref{tab:use_tasks}) that are common to Instagram, which cover a wide range of use common to many mobile applications (e.g., reading text, viewing pictures, swipes, touch, text input). 
Video recordings were made of the session, and coded for the analysis.


\begin{table}[tbh]
  \caption{Common use scenarios in Instagram framed as design tasks for Study 2}
  \label{tab:use_tasks}
  \Description[Short description]{Long description}
  \begin{tabular}{c|l}
    \toprule
    Task & Description\\
    \midrule
    1 & Like and bookmark a post\\
    2 & Read a post and post a comment\\
    3 & Send a message to a friend about where you are\\
  \bottomrule
\end{tabular}
\end{table}



\subsection{Study 2: Designers using Augmented Virtuality Impairment Simulation}

\subsubsection{Participants}
We recruited 10 (5 female) experienced HTML-CSS-JS designer-developers, ages 21-31 (mean 24.4) years. 
The participants were selected to not have any pre-existing uncorrected visual impairments--for example, myopia corrected by spectacles was allowed, but colour blindness was not. 
To ensure participants had the ability to perform the study task, we administered a coding test for HTML-CSS-JS, finding that all participants had the requisite skills (9 of 10 scored similarly, with only one participant demonstrating greater coding ability).
The participants were divided into two conditions. 
In the first condition, C1, participants were given \name (see Table~\ref{tab:empathd_hardware} for the hardware used to run Empath-D), and in the second, C2, they were given \textit{Google Accessibility Scanner}~\cite{noauthor_accessibility_nodate}, a diagnostic tool for Android mobile app accessibility. 
To further ensure that participants in each group were unbiased in ability, we administered the Need For Cognition~\cite{cacioppo_need_1982} scale that reflects an individual's inclination towards effortful cognitive abilities (in this case, redesign). 
We did not find significant differences in the mean scores between the two groups of users (C1: 10.6 (SD = 5.9) and C2: 13.8 (SD = 7.8))).

\subsubsection{Method}
Participants came in for a full day study and were asked to perform the role of a mobile app designer, and redesign a mockup of Instagram for cataract impaired users to support the same three use scenarios described in S1 and for the same reasons (see Table~\ref{tab:use_tasks}). 
This allows us to match the usability challenges identified by cataract impaired users to the usability challenges addressed by designer-developers. 
For each task, participants were given 100 mins to redesign and develop the mockup, rolling over from each session to finally end with a working prototype.
Participants were free to choose their process of working with the tools provided. 
They could modify the designs as they pleased (e.g., reposition, resize, recolour, remove), with only one limiting condition: they should not take away functions from the user interface. 
For instance, if the bookmark button was the sole means by which to bookmark a post, users cannot remove the bookmark button without implementing bookmarking in a different manner.

\begin{table}[tbh]
	\caption{Hardware used to run Empath-D.}
	\label{tab:empathd_hardware}
	\Description[Short description]{Long description}
	\resizebox{\columnwidth}{!}{
		\begin{tabular}{ll}
			\toprule
			Component & Device Used\\
			\midrule
			VR display & Samsung Galaxy S7\\
			RGBD camera & Intel RealSense SR300\\
			Computer & CPU: Intel Core i7-3720QM (4 cores, 2.6 GHz)\\
			 & RAM: 16 GB DDR3\\
			 & GPU: NVIDIA GeForce GT 640M\\
			Physical IO & Samsung Galaxy S7\\
			smartphone & \\
			\bottomrule
		\end{tabular}
	}
\end{table}

\begin{figure}
\includegraphics[width=0.95\columnwidth]{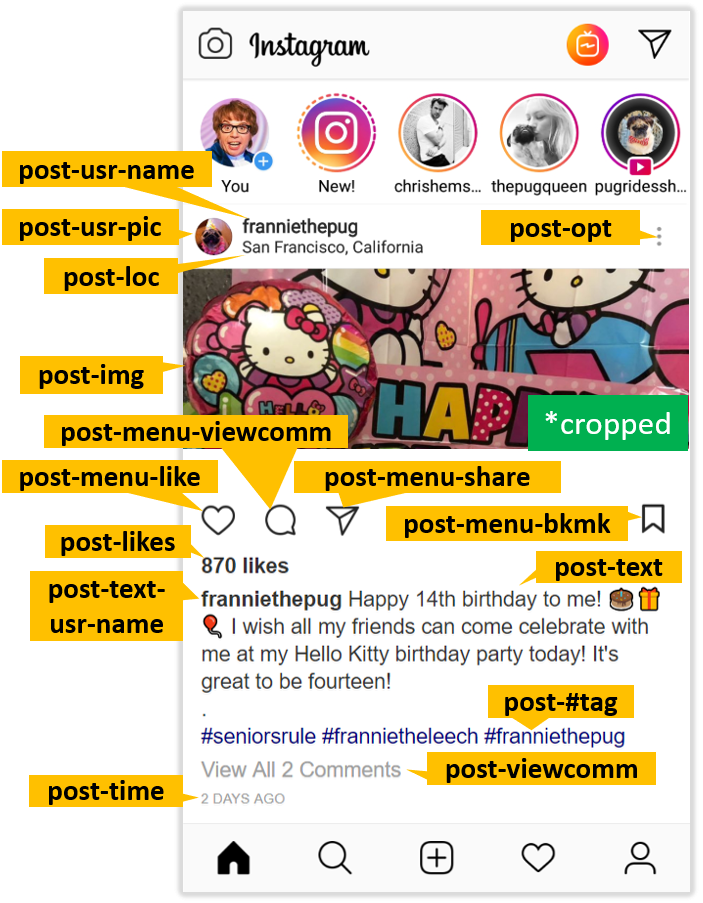}
\caption{First page of the mockup baseline design used for S2 with example labels of UI elements used in the analysis. \textit{Image source: Instagram, \copyright~franniethepug}}
\label{fig:post_page_labelled}
\Description[A snapshot of Instagram annotated with a partial set of the labels of the UI elements used in the mockup of Instagram.]{A snapshot of Instagram annotated with a partial set of the labels of the UI elements used in the mockup of Instagram.}
\end{figure}

We developed the mockup of Instagram using HTML-CSS-JS, and set it up to be compiled using Cordova~\cite{noauthor_apache_nodate} into an Android apk file.
Only two popular JavaScript-based APIs were utilised to minimise the learning that was required of participants.
\textit{jQuery Mobile} was used to provide a mobile-like experience, yet having all the code in a single HTML file to simplify the organisation of the code.
\textit{jQuery} was provided to support simpler coding of the UI logic.
This served as the baseline application mockup (see Figure~\ref{fig:post_page_labelled}) that all participants started with.

We deliberately chose HTML-CSS-JS as the underlying means to develop this app as it provides three distinct advantages: 1) users can focus on aspects of design, as compared to native-app coding, 2) they may easily inspect and simulate changes using browser-based mobile emulation (e.g., in Google Chrome), and 3) the HTML-CSS-JS can be run through existing accessibility checkers for WCAG 2.0 compliance.

Participants were briefed on the WCAG 2.0~\cite{caldwell_web_2008}, and given a link to an abstract version of the WCAG 2.0 geared towards older users (WCAG 2.0-Elderly)~\cite{arch_developing_2010}. 
The WCAG 2.0-Elderly was provided as a guiding document since it provides the closest match to our desired elderly demographic of which 78.6\% suffer from cataracts~\cite{ho1997eye}. 
Participants were also given a description of cataracts, and its underlying symptoms~\cite{noauthor_cataracts_2018}.
They also had unfettered access to the Internet to support seeking information on any aspect of the task.
Lastly, to help participants to lower the challenges of implementation and focus them on design, we provided them with: 1) cheatsheets for HTML-CSS-JS, and 2) assistance from the experimenters on implementation (e.g., how to adjust HTML elements, code a JS function, or image editing) but not design issues. 
The support provided was meant to reduce the impact of individual differences in coding ability, which may negatively affect participants with weaker coding ability by distracting them from focusing on design.

All participants performed the study using a 13-inch Macbook, installed with tools common to web development (e.g., Sublime Text~\cite{noauthor_sublime_nodate} / Atom, with code completion / syntax highlighting features). 
They were also free to install any tools that they preferred to use, though no participants felt the need to and were comfortable with the tools given. 
An external monitor was connected to the computer and was recorded using a video camera to capture holistic data of all that the user was doing in the design task (e.g., searching for information, focusing on an element for redesign). 
A separate video camera in parallel was also used to capture the interactions that a user had with \name or Google Accessibility Scanner. 
We adopted a think-aloud protocol, getting participants to verbalise their thoughts as they performed the design task. 
At the end of each task, participants were asked to fill in the NASA Task Load Index (NASA-TLX) with reference to the whole design task and System Usability Scale (SUS) with reference to the system they used (C1: \name or C2: Google Accessibility Scanner).
\section{Results}

\subsection{S1: Cataract Impaired Users}
\begin{figure}[tbh]
\includegraphics[width=0.95\columnwidth]{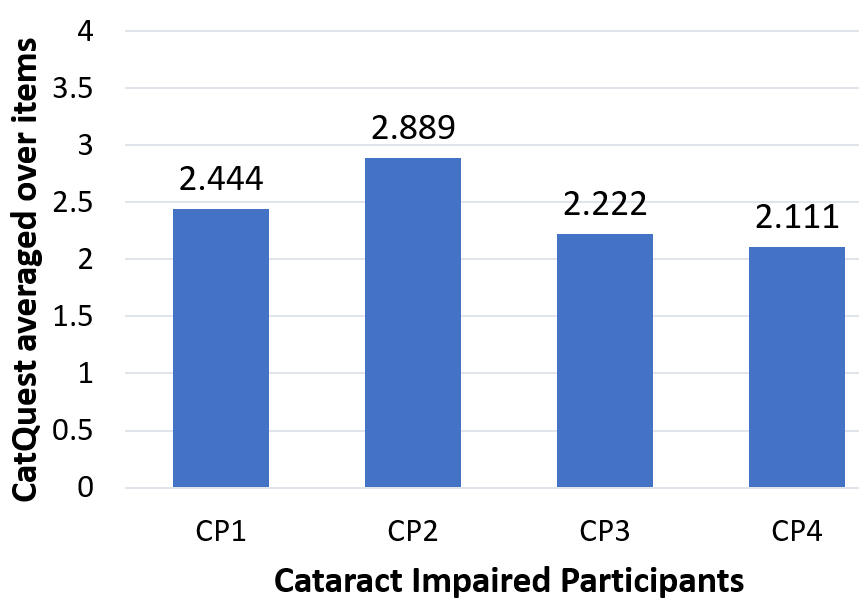}
\caption{All participants had similar mild-moderate (between 2: ``some difficulty'' to 3: ``great difficulty'') levels of cataract impairments.}
\label{fig:catquest_scores}
\Description[A bar graph showing that the averaged CatQuest-9SF scores of the cataract impaired participants are close,]{A bar graph showing that the averaged CatQuest-9SF scores of the cataract impaired participants are close, and fall between some difficulty (2) to great difficulty (3)}
\end{figure}

From the CatQuest-9SF, all participants were found to match our targeted mild to moderate level of impairments, with cataract impairment scores ranging from 2.111-2.889 (2 being ``some difficulty'' and 3 being ``great difficulty'') (See Figure~\ref{fig:catquest_scores}).


We analysed the observations in S1 to establish a base set of usability challenges that cataract impaired users face (See Cataract Impaired Users, Table~\ref{tab:usability_challenges}), with respect to the core use tasks detailed in Table~\ref{tab:use_tasks}. 

\begin{table}[tbh]
  \caption{Usability Challenges identified in C1 and C2, mapped to WCAG 2.0-Elderly and what Impaired Users identified.}
  \label{tab:usability_challenges}
  \resizebox{\columnwidth}{!}{
  \begin{tabular}{l|cc|cc}
    \toprule
    Usability & WCAG 2.0 & Cataract & C1 & C2\\
    Challenges & -Elderly & Impaired Users & (\name) & (GAS)\\
    \midrule
    Font size & \checkmark & \checkmark & \checkmark & \checkmark\\
    Letter spacing &  & \checkmark & \checkmark & \\
    Contrast & \checkmark & \checkmark & \checkmark & \checkmark\\
    Image visibility & \checkmark & \checkmark & \checkmark & \checkmark\\
    Icon visibility & \checkmark & \checkmark & \checkmark & \checkmark\\
    Hit Target & & & \checkmark & \checkmark\\
  \bottomrule
  \end{tabular}
  }
\end{table}

The challenges that they faced match the WCAG 2.0-Elderly guidelines in all challenges except \textit{letter spacing}. 
One participant noted that bold text (e.g., \textit{post-text-usr-name}) was harder to see than non-bold text (e.g., \textit{post-text}), as the lower ratios of letter to stroke spacing caused letters to ``clump together'' (See Figure~\ref{fig:post_page_labelled} for italicised examples). 
The converse problem also occurred, with low font weights, two users found text hard to read. 
All users noted that with larger font sizes, text became more readable. 
This demonstrates the need to \textit{appropriately balance font size, font weight and letter spacing}--a design requirement that is often left to the judgement of designers.
All participants also had problems reading low contrast grey text such as those of \textit{post-viewcomm} and \textit{post-time} (See Figure~\ref{fig:post_page_labelled}). 
For all participants, \textit{contrast} was the biggest problem, often resulting in the inability to perceive the content at all. \textit{post-viewcomm} and \textit{post-time} are deliberately designed to be grey, such as to convey its relatively lower importance in the UI. 
However, with elderly users, this results in the loss of perceiving some information about the post. 
For \textit{post-time}, elderly users cannot tell when the post was posted.
For \textit{post-viewcomm}, elderly users cannot perceive how much discussion is going on and consequently decide if to click and find out more.


\subsection{S2: Identifying and Fixing Design Problems}

We performed a detailed breakdown of the changes made to the base mock-up at the element level, and separated them into positive and negative categories for each condition (see Figure~\ref{fig:coverage_matrix}). 
We determined positive/negative changes by how they are in line with the WCAG 2.0 and data from cataract impaired users (S1), but ignore for the magnitude of changes and how they affect overall usability.
For example, if a design guideline indicates that font sizes should be enlargeable/enlarged to support vision, and the user made a change to enlarge the font size, this counts as a positive change. 
However, if the user instead reduces the font size, this registers as a negative change.
We further separated each positive and negative category into the key usability challenges (e.g., font size, letter spacing).

\subsubsection{Accuracy and Coverage}
\begin{figure}[tbh]
\includegraphics[width=0.49\textwidth]{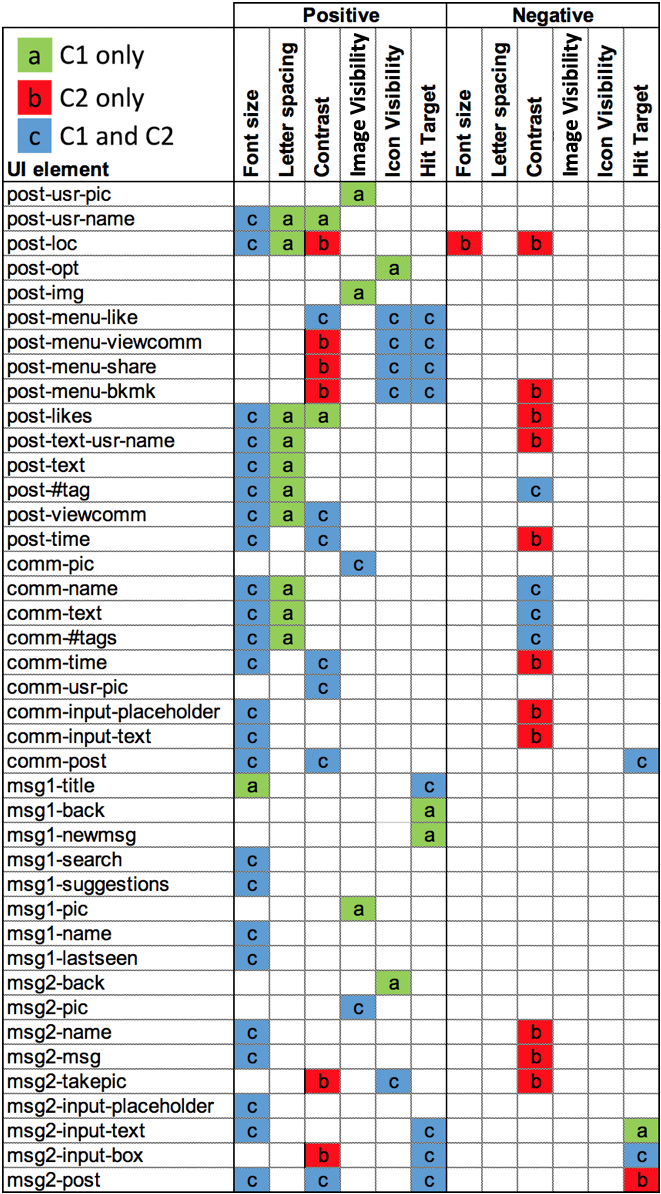}
\caption{Coverage matrix of the usability challenges positively/negatively addressed by designer-developers. \colorbox[RGB]{159,205,98}{a} indicates challenges identified only by Empath-D (C1). \colorbox[RGB]{236,50,36}{b} indicates challenges identified only by Google Accessibility Scanner (C2). \colorbox[RGB]{103,155,207}{c} indicates challenges identified by \textit{both} Empath-D and Google Accessibility Scanner. Blank cells indicate that no users identified challenges in either condition for that UI element.}
\label{fig:coverage_matrix}
\Description[The coverage matrix of the usability challenges positively or negatively addressed by designer-developers. The challenges are grouped by UI element, positive and negative categories, and by usability challenge.]{The coverage matrix of the usability challenges positively or negatively addressed by designer-developers. The challenges are grouped by UI element, positive and negative categories, and by usability challenge.}
\end{figure}

In C1 (\name), participants were able to more accurately (94.2\%, 180 positive changes) identify usability challenges than without (C2: Google Accessibility Scanner; 85.6\%, 160 positive changes). 
From Figure~\ref{fig:coverage_matrix}, we see that \name supports positively identifying 19.4\% more usability challenges across the different UI elements. 
Notably, \name allowed participants to identify \textit{letter spacing} problems, a usability challenge not picked up at all in C2. 
This was because Google Accessibility Scanner did not report issues of letter spacing. 
Users in C1 however, could \textit{directly observe} through \name impairment simulation that words ``clumped together'' (P4), and made appropriate changes.
Participants in C2 picked up 40\% more contrast usability challenges than in C1, however, this is at the detriment of wrongly identifying 100\% more contrast usability challenges across the different UI elements. 
This indicates that Google Accessibility Scanner (C2) is unreliable in identifying contrast usability challenges. 

Examining for all UI elements where positive changes were identified, participants in C1 were able to uniquely identify usability challenges in  7 (15.2\% of all elements) UI elements: \textit{post-usr-pic}, \textit{post-opt}, \textit{post-img}, \textit{msg1-back}, \textit{msg1-newmsg}, \textit{msg1-pic}, and \textit{msg2-back}. 
This is pertinent particularly since images (\textit{post-img}) (see Figure~\ref{fig:post_page_labelled}), are a central mechanic to deciding to further interact in Instagram, and that cataract impaired users (in S1) reported not being able to perceive the details in images, particularly those that are complex (e.g., colourful images with fine details that may obscure the focal subject of the image).

\subsubsection{Magnitude of Changes}
To better understand the effects of the changes made, we collated the UI elements into their functional categories and analysed the differences between C1 and C2 (see Table~\ref{tab:ui_function_changes}). 
Only letter spacing is definitively significant as users in C2 did not make letter-spacing changes at all. 
While no significant differences were found for the rest of the conditions, this is largely due to the nature of a design task. 
Take for instance the size of text. The only change that may be specified is that an element should meet a minimum size in order to be perceivable. 
However, users may increase the size of text much more than that, as they feel that it does not impact scrolling significantly.

Given that text is central to performing T1-T3, we see that in both C1 and C2, participants increased the font sizes, with C1 being larger across all UI functions. 
Letter spacing was modified only in C1, with spacing ranging from 1.5-1.83 pixels across the different UI functions.

This was corroborated by the observational data captured during the experiments. 
In C2, users found it hard to understand and use the relevant WCAG 2.0 success criteria (1.4.4 - Resize text). 
While 1.4.4 suggests that ``text can be resized without assistive technology up to 200 percent without loss of content or functionality'', participants also considered the resulting penalty to interaction (having to scroll a lot more due to larger text/buttons).  
Conversely, participants in C1 had the ability to directly simulate the cataract impairment using \names, verifying that the font sizes that they chose were sufficiently large such that they could be perceived, and yet minimise the penalty to interaction.

\begin{figure*}[tbh]
\includegraphics[width=0.7\textwidth]{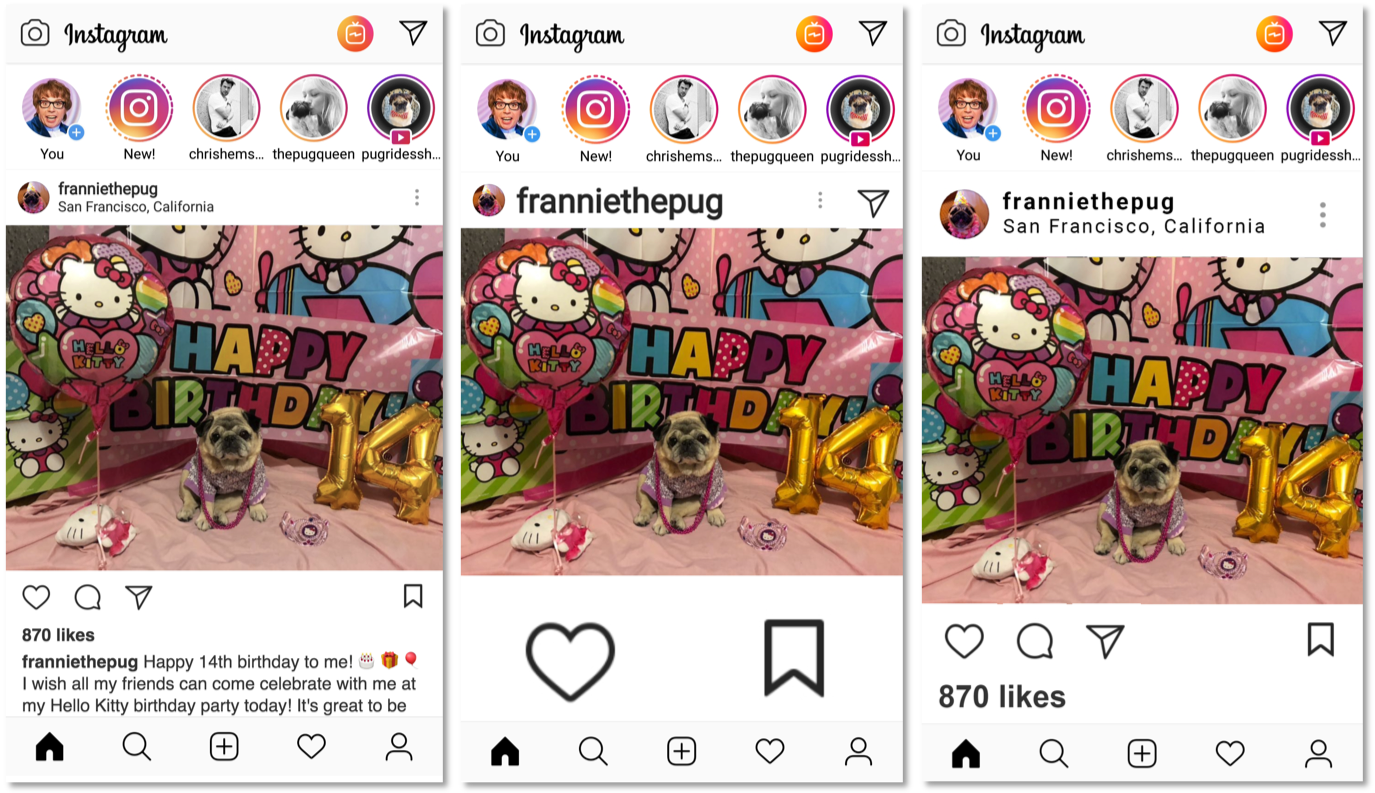}
\caption{ Balancing interaction design. Left: Unmodified base app. Centre: Design from P8 (C2: Google Accessibility Scanner) with very large like and bookmark buttons sacrificing content and interaction. Right: Design from P10 (C1: \names) showing balance in design despite changes made to increase elements. \textit{Image source: Instagram, \copyright~franniethepug}}
\label{fig:balancing_interactiondesign}
\Description[Three screenshots of the Instagram mockup are shown left to right. The first shows the baseline unmodified mockup. The second image shows a design from a participant in Condition 2, Google Accessibility Scanner, with large like and bookmark buttons that sacrifice content and interaction. The third image shows a design from a participant in Condition 1, Empath-D, with moderately larger buttons and text, showing balance in design without sacrificing the ability to observe more in a single page view.]{Three screenshots of the Instagram mockup are shown left to right. The first shows the baseline unmodified mockup. The second image shows a design from a participant in Condition 2, Google Accessibility Scanner, with large like and bookmark buttons that sacrifice content and interaction. The third image shows a design from a participant in Condition 1, Empath-D, with moderately larger buttons and text, showing balance in design without sacrificing the ability to observe more in a single page view.}
\end{figure*}

Figure~\ref{fig:balancing_interactiondesign} shows the baseline design (left), an example design by P8 in C2 (centre), and P10 from C1 (right). 
P8 focused on supporting the design task T1, enlarging the two buttons of like and bookmark to 11 times larger a hit and visible area than the baseline. 
This resulted in less content being able to be seen in one page view (pushing down \textit{post-likes}).
While the much larger buttons help users to accurately press the buttons, we find from S1 that the cataract impaired users reported not having problems with pressing the buttons due to their familiarity with using mobile apps such as WhatsApp or Facebook.
Comparatively, P10's design made moderate icon size, font size, and letter spacing adjustments, retaining more content that may be observed in one page view.

\begin{table}[tbh]
  \caption{UI elements, grouped by key functions showing mean changes made by designer-developers.}
  \label{tab:ui_function_changes}
  \resizebox{\columnwidth}{!}{
  \begin{tabular}{lcccccc}
    \toprule
     & & Font & Letter &  Cont & Visi & Hit\\
    UI & Cond & Size & Spacing & -rast & -bility & Target\\
    Function & -tion & (\%) & (px) & (\%) & (\%) & (\%)\\
    \midrule
    button  & 1 & 58 & 1.83 & 18 & 98 & 118\\
            & 2 & 44 & 0 & 60 & 185 & 107\\
    image-content   & 1 & - & - & - & 5 & -\\
                    & 2 & - & - & - & 0 & -\\
    image-profile   & 1 & - & - & - & 63 & -\\
                    & 2 & - & - & - & 16 & -\\
    text-content    & 1 & 52 & 1.5 & -6 & - & -\\
                    & 2 & 31 & 0 & -5 & - & -\\
    text-input  & 1 & 59 & - & 0 & - & 23\\
                & 2 & 34 & - & 0 & - & 20\\
    text-navigation & 1 & 28 & - & 0 & - & -\\
                    & 2 & 25 & - & -1 & - & -\\
    text-status     & 1 & 43 & 1.63 & 31 & - & -\\
                    & 2 & 28 & 0 & 12 & - & -\\
    text-username   & 1 & 68 & 1.69 & 18 & - & -\\
                    & 2 & 40 & 0 & 8 & - & -\\
  \bottomrule
  \end{tabular}
  }
\end{table}


\subsection{Usability of Tools}

\begin{figure}[tbh]
\includegraphics[width=0.95\columnwidth]{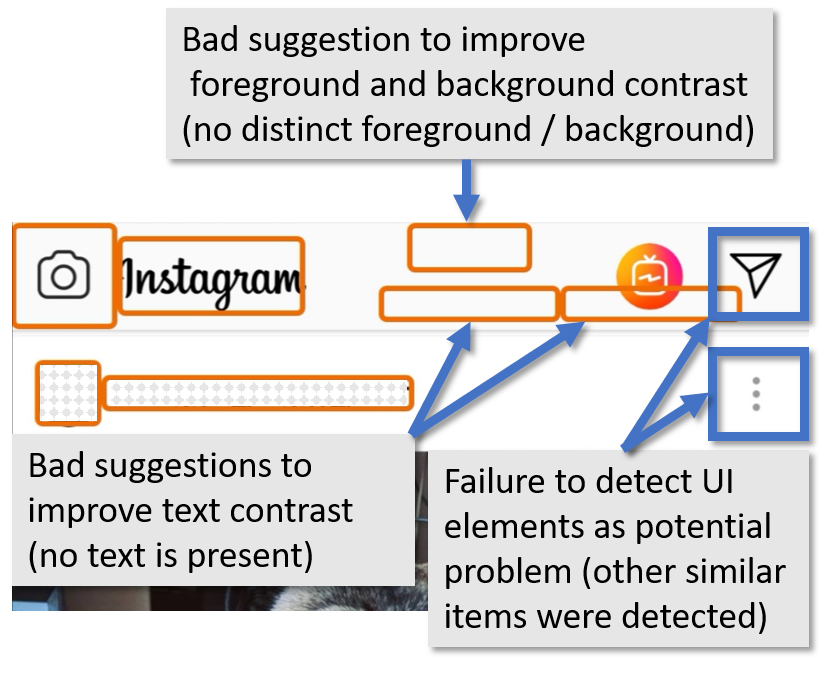}
\caption{Scan from Google Accessibility Scanner showing obvious bad suggestions. Google Accessibility Scanner flags out potentially problematic areas with orange boxes.}
\label{fig:badGAS}
\Description[A partial screenshot of the Instagram mockup with orange boxed annotations from Google Accessibility scanner.]{A partial screenshot of the Instagram mockup with orange boxed annotations from Google Accessibility scanner. Google Accessibility Scanner provided some obvious bad suggestions.}
\end{figure}

Participants using \name reported higher SUS scores of 76.3 (\textit{Acceptable}) as compared to using Google Accessibility Scanner, having a score of 65.2 (\textit{Marginal})~\cite{bangor_determining_2009}. 
One contributing reason is that all users found that Google Accessibility Scanner can be unreliable. 
It reported suggestions in app designs despite its obvious irrelevance or was unable to flag issues with some UI elements despite similar UI elements being flagged elsewhere in the UI, e.g., \textit{post-menu-share}) (See Figure~\ref{fig:badGAS}).


However, 2 users (40\%) indicated that overall Google Accessibility Scanner was useful for the in situ examination of the accessibility of mobile apps, providing a more concrete means (compared to WCAG guidelines) to identify potential problems that users may then map to cataract impairment. 

All users in C1 on the other hand found that \name provided a concrete means to immediately identify usability problems. 
One user noted that \name enabled her to identify problems with letter-spacing, which she stated was not in the WCAG 2.0. 
However, this is inaccurate--the WCAG 2.0 does contain recommendations about adjusting letter spacing. 
We believe this reflects two issues in the WCAG 2.0: 1) that the WCAG 2.0 while comprehensive, is onerous for use, and 2) the difficulty in mapping the pathophysiology of impairments to specific guidelines. The interviews with users alluded to this. Users found the WCAG 2.0 to be extremely lengthy (4, 40\%), and often found it hard to map guidelines relevant to cataract impairment (5, 50\%). However, they noted that the WCAG 2.0-Elderly (and WCAG 2.0 itself) is a rich source of information to provide conceptual guidance on possible issues to focus on (8, 80\%).

No significant differences were found in the scores of the NASA-TLX.

\begin{table}[tbh]
  \caption{System Usability Score and NASA-TLX}
  \label{tab:sus_tlx}
  \begin{tabular}{lcc}
    \toprule
    Condition & SUS & NASA-TLX\\
    \midrule
    1 & $76.33^{\ddag}$  & 11.64\\
    2 & $65.17^{\dag}$ & 10.71\\
    \bottomrule
    \multicolumn{3}{r}{not Acceptable / $Marginal^{\dag}$ / $Acceptable^{\ddag}$~\cite{bangor_determining_2009}}\\
  \end{tabular}
\end{table}

\subsubsection{Automated Accessibility Checking}

We ran an automated accessibility checking tool \textit{aChecker}~\cite{gay_achecker:_2010} on the final app designs from S2 over WCAG 2.0 guidelines, and at the \textit{AA} level of compliance. 
The W3C does not recommend \textit{AAA} conformance as a general policy for entire sites as it is not possible to satisfy all the success criteria for some content. 
\textit{aChecker} bins checks into 3 categories of \textit{Known}, \textit{Likely}, and \textit{Potential} problems.

\begin{figure}[tbh]
\includegraphics[width=0.95\columnwidth]{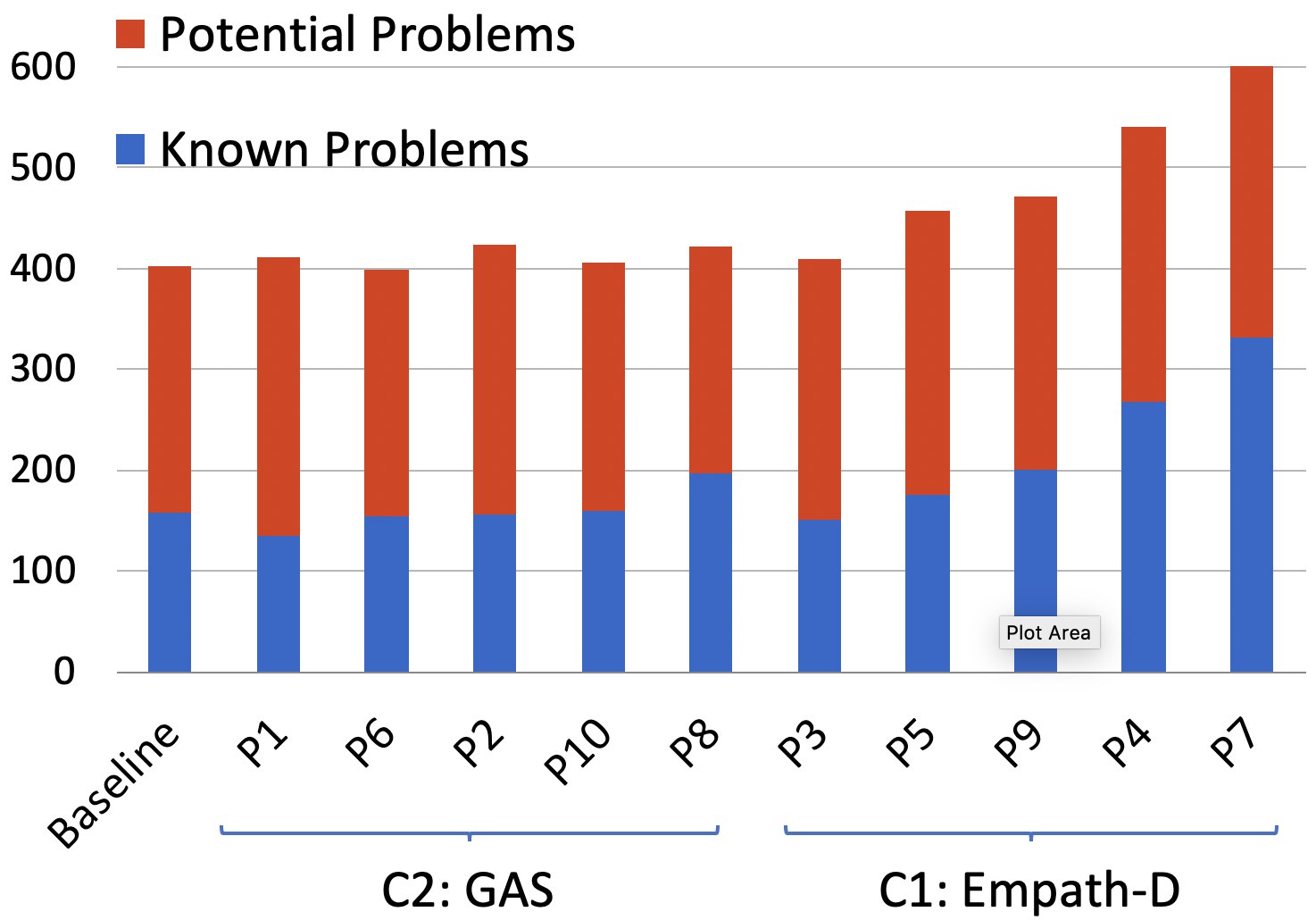}
\caption{More known and potential accessibility problems identified in C1 than in C2 when aChecker is used.}
\label{fig:achecker_accessibility_check}
\Description[A bar chart that shows the known and potential accessibility problems identified using aChecker in the baseline Instagram mockup, participants in C1, and C2. More known and potential accessibility problems are identified in C1 and C2]{A bar chart that shows the known and potential accessibility problems identified using aChecker in the baseline Instagram mockup, participants in C1, and C2. More known and potential accessibility problems are identified in C1 and C2}
\end{figure}

\textit{aChecker} flagged $40.5\%$ more \textit{Known} and $7.4\%$ more \textit{Potential} problems in C1 than in C2 (See Figure~\ref{fig:achecker_accessibility_check}). 
No significant differences were found when comparing C1 and C2 in both \textit{Known} and \textit{Potential} conditions.
From Figure~\ref{fig:coverage_matrix}, we know that C1 has better accuracy at identifying and fixing designs for cataract impaired users. 
We examined the issues that were flagged by \textit{aChecker}, and found that many of the identified issues related to missing alt text labels for images. 
Such a requirement is only necessary to support the use of general accessibility functions (e.g., talk back to read out what an element is). 
However, our focus is on designing specifically for an impairment, with no reliance on OS-provided accessibility support.
The results from \textit{aChecker} demonstrate the limitations of automated accessibility checking, which are inaccurate when used for designing  specific impairments. 

\section{Discussion}

\ken{REBUTTAL: R1 asked about the relevance of our work to the CHI readership. While developers can directly benefit from the proposed tool, our work can has a broader impact for the CHI accessibility community and potentially many impaired users. We believe this work can motivate more in the CHI accessibility community to consider doing empirical work that will build up a wide base of impairments, and therefore impact more users. This work is a cornerstone in a whole field of work that can explore different impairments and combinations of impairments; all experienced in the environmental and interaction contexts provided by augmented virtuality simulation such as Empath-D [18] to support mobile app design. We will clarify this in our discussion in the revision.}

\ken{REBUTTAL: We agree with R2 that the current study results can be applied to similar scenarios. We believe our work demonstrates the efficacy of our methodology and the relevance of augmented virtuality impairment simulation for design. As discussed in relation to the above para on relevance of work, this sets the means by which more empirical studies may be conducted for different impairments, establish empirical grounds for different types of impairment simulations, and potentially generate a library of impairments that may be open-sourced. We agree with R2 that larger sample sizes would help, but leave that for future work in different types of impairments and/or environmental contexts. We will include this in our discussion in the revision.}

\subsubsection{The need for more and better tools to design mobile apps for accessibility}
While much has been explored in designing \textit{web content} for general accessibility, tools to support examining and designing mobile apps for accessibility largely remain an impoverished space.
They primarily rely on the same basis that web content relies on--design guidelines--which have been previously suggested in previous studies to be insufficient to ensure web accessibility ~\cite{clegg2014investigating, mankoff2005your, milne_are_2005}. 
Our study utilising Google Accessibility Scanner reinforces this position, demonstrating its mismatch with user concerns as well as its potential for false positives. Users despite their best efforts, found it hard to map WCAG guidelines to cataract impairments, and even when they did, had to guess the effects of adjustments to the design.

While \name allowed users to make adjustments to their designs concretely, it suffers from several problems. 
Firstly, the \textit{attention} of users is required for them to examine usability problems. 
Without systematically examining the various components in a user interface, users may fixate on a particular issue, and miss out on identifying others. 
Secondly, users often want to find \textit{several usability problems} at once when using \names. 
However, being immersed in the VR, they are unable to make a note of all the identified issues.
One potential solution to these two problems is to support annotation in the VR. 
A cursor overlay would allow users to use digital styluses to circle and note down problematic aspects of the UI. 
These annotations (e.g., the circular strokes), may be used as a means to develop a heat map over areas that have been considered and remind users to examine other aspects of the UI.

\subsubsection{Complementing augmented virtuality evaluation and guidelines}
Even when users focus their attention to inspect a suspected problem area of the UI, they may not be able to conceptualise all problems related to that element. 
For instance, a user may observe that an icon button has insufficient stroke width to be adequately perceived, and adjusts accordingly. 
However, an alternate solution to this is to provide text labels, which may be a more appropriate solution for elderly users since it states its function, rather than relying on a user's memory.
\textit{Impairment-focused accessibility scanning} may be one potential solution. A filtered set of guidelines (possibly stemming from the WCAG), may be employed to scan the UI elements in the app to flag potential issues. 
This marries the benefits of \textit{concrete experience} provided by \name with the \textit{completeness} provided by guidelines.

\subsubsection{Simulation fidelity and range of testing}
\name needs to be further developed to support greater sensing capabilities that match or even supersede those provided by existing smartphone capabilities. 
The sensing capabilities of the physical smartphone should be appropriately represented in the VR in a seamless fashion. 
For instance, 1) location in the virtual environment can be mapped to the virtual coordinates in the environment and reflected in the virtual phone, and 2) virtual cameras show frames of the virtual environment on the virtual smartphone. 
These are but two examples of enhancements that support existing geolocation and camera applications respectively. 
By providing capabilities that may not be commonly available on existing hardware, and by providing realistic environmental simulations such as the home or the streets, \name becomes a more powerful tool for interaction design.
For example, by developing a virtual depth camera into \names, one may test the interaction concepts in potentially dangerous environments such as the streets, which may present greater danger to users with disabilities.
Naturally, the use of \name also goes beyond designing for users with disabilities--absent impairment simulation--\name still provides naturalistic interaction in a VR.
This opens many opportunities for research that deals with novel interaction concepts with mobile and wearable devices.

\subsubsection{Towards a Library of Impairments}
Impairments are diverse in type, severity, and presentation.
The studies in this paper are a first step in establishing a \textit{library of impairments}, focusing in particular on cataract visual impairments.
The envisioned library of impairments would allow designers to choose a representative group of users with disabilities to design for.
This in particular would help the issues of a lack of training and confusion with general accessibility guidelines as highlighted by Lazar et al.~\cite{lazar_improving_2004}.
However, research involving users with accessibility needs are often fraught with difficulty (e.g., representing users in experimentation due to diversity, low availability)~\cite{sears_representing_2011, newell_user-sensitive_2011}.
We call on the accessibility research community to support this vision to develop a library of impairments.

\section{Conclusion}

In this paper, we studied the effects of using \name, an \textit{augmented virtuality} impairment simulation system, to support experienced designer-developers to design a mobile application for cataract-impaired users. 
Our studies with 4 cataract-impaired elderly and 10 experienced mobile developers show that with the aid of augmented virtuality and WCAG 2.0 guidelines, the developers were better able to identify the usability challenges of cataract-impaired users and made more positive changes to the app designs as compared with using Google Accessibility Scanner with WCAG 2.0 guidelines. 
The developers also noted that \name is more usable than Google Accessibility Scanner.
\name helps them to immediately and uniquely identify usability problems (i.e., letter spacing) and see the effect of changes made, while the accessibility scanner often points out problems irrelevant to the target users and can be unreliable for many design components of the target application. 
\begin{acks}
This research is supported by the National Research Foundation, Prime Minister's Office, Singapore under its IDM Futures Funding Initiative.
\end{acks}

\bibliographystyle{ACM-Reference-Format}
\balance
\bibliography{bibliography}

\end{document}